\documentclass[twocolumn,showpacs,preprintnumbers,amsmath,amssymb,nofootinbib]{revtex4-1}
\usepackage{graphicx}% Include figure files
\usepackage{dcolumn}% Align table columns on decimal point
\usepackage{bm}% bold math
\usepackage{natbib}

\usepackage{amssymb}
\usepackage{amsfonts}
\usepackage{amsmath}
\usepackage{bm}

\usepackage{float}

\usepackage{color}

\definecolor{nv}{rgb}{0.1,0.1,0.6}
\definecolor{pr}{rgb}{0.2,0.1,0.5}
\definecolor{mg}{rgb}{0.4,0.0,0.4}

\newcommand{\pareq}{\stackrel{\textrm{par}}{=}}

\newcommand{\beq}{\begin{equation}}
\newcommand{\eeq}{\end{equation}}
\newcommand{\beqy}{\begin{eqnarray}}
\newcommand{\eeqy}{\end{eqnarray}}
\newcommand{\beqyn}{\begin{eqnarray*}}
\newcommand{\eeqyn}{\end{eqnarray*}}
\newcommand{\nl}{\newline}

\newcommand{\bs}{\begin{slide}}
\newcommand{\es}{\end{slide}}
\newcommand{\bc}{\begin{center}}
\newcommand{\ec}{\end{center}}
\newcommand{\bmin}{\begin{minipage}}
\newcommand{\emin}{\end{minipage}}

\newcommand{\bi}{\begin{itemize}}
\newcommand{\ei}{\end{itemize}}

\usepackage{latexsym}

\usepackage{dsfont}
\usepackage{multirow}

\newcommand{\bea}{\begin{eqnarray}}
\newcommand{\eea}{\end{eqnarray}}
\newcommand{\be}{\begin{equation}}
\newcommand{\ee}{\end{equation}}

\newcommand{\ud}{\mathrm{d}}

\newlength\savedwidth

\newcommand{\uvec}[1]{\boldsymbol{#1}}

\newcommand{\pure}{\text{pure}}
\newcommand{\phys}{\text{phys}}

\begin{document}

\preprint{APS/123-QED}

\title{Measurement of Optical  Orbital and Spin Angular Momentum: Implications for Photon Angular Momentum}% Force line breaks with \\

\author{Elliot Leader}
 \affiliation{Blackett Laboratory\\Imperial College London\\Prince Consort Road\\London SW7 2AZ, UK}
 \email{e.leader@imperial.ac.uk}
%\affiliation{%

% \affiliation{

\date{\today}% It is always \today, today,

\begin{abstract}
The expression for the total angular momentum carried by a laser optical vortex beam, splits, in the paraxial approximation, into two terms which seem to  represent orbital and spin angular momentum respectively. There are, however, two very different competing versions of the formula for the spin angular momentum, one based on the use of the Poynting vector, as in classical electrodynamics, the other related to the canonical expression for the angular momentum which occurs in Quantum Electrodynamic. I analyze the possibility that a sufficiently sensitive optical measurement could decide which of these corresponds to the actual physical angular momentum carried by the beam.
\end{abstract}

\pacs{42.50.Wk, 42.50.Tx, 42.25.Bs, 11.15.-q, 12.20.-m, 14.70.Bk}% PACS, the Physics and Astronomy
\maketitle

\section{\label{sec:1}}
Quantum Electrodynamics (QED) textbooks, for over half a century, have stressed that the total angular momentum of a photon cannot be split into  spin and orbital angular momentum (OAM) parts \emph{in a gauge invariant way}. Hence the extraordinary reaction, (for reviews see \cite{Leader:2013jra,Wakamatsu:2014zza}) a few years ago, when Chen et. al. \cite{Chen:2008ag} produced what they claimed, was precisely such a gauge invariant split. They introduced fields which they called $\uvec A_\pure$ and $\uvec A_\phys$, but which  are identical to the  fields  in the Helmholz decomposition into longitudinal ($\bm{A}_\| $) and transverse ($\bm{A}_\bot $)     components with
\beq \label{perppar}  \uvec\nabla\times\uvec A_\|=\uvec 0, \quad \textrm{and} \quad \uvec\nabla\cdot\uvec A_\bot=0 \eeq
and obtained
\beq \label{CHEN}
\uvec J=\underbrace{\int\ud^3x\,\uvec E\times\uvec A_\bot}_{\bm{S}}+\underbrace{\int\ud^3x\,E^i(\uvec x\times\uvec\nabla) A^i_\bot}_{\bm{L}} \eeq
and since $ \uvec A_\bot $ and $\bm{E} $ are unaffected by gauge transformations, they appeared to have achieved the impossible.
But exactly the same expression, Eq.~(\ref{CHEN}), had  already been given in the textbook of Cohen-Tannoudji et.al. \cite{CohenTannoudji:1987bi} in 1987 (!), and   some years after that van Enk and Nienhuis \cite{vanEnk1994}  had pointed out that, actually,  the split was a failure because the spin and OAM  operators did not satisfy correct angular momentum (AM) commutation relations, i.e. they showed that
 \beq \label{vENcom} [\, S^{ i }\, , \, S^{ j} \,] =0 \, \qquad \textrm{and} \qquad [\, \bm{L} \,,\,\bm{S} \,] \neq 0 .\eeq
 Thus the claim of Chen et. al. is unquestionably incorrect.\nl
 Despite the fact that the  operators $\bm{S}$ and $\bm{L}$ in Eq.~(\ref{CHEN}) are not genuine AM operators, we shall see that they play an important role in laser optics. In the following,
 because of the complicated history involved, and because the expression Eq.~(\ref{CHEN}) closely resembles the usual \emph{canonical} expression for photon angular momentum (which simply has $\bm{A}_\bot$ replaced by $\bm{A}$) I shall refer to it   as the \emph{gauge invariant canonical}  (gican) version of the AM. \footnote{often abbreviated to gic}. Thus
  \beq \label{Jgican} \uvec J_{\text{gican}}= \int\ud^3x\,\uvec j_{\textrm{gican}} \eeq
  where the total angular momentum \emph{density} is
  \beq \label{jgican} \uvec j_{\textrm{gican}}(x)= \uvec l_{\textrm{gican}}(x) + \uvec s_{\textrm{gican}}(x) \eeq
  and where the spin and orbital densities are
  \beq \label{sandlgican} \uvec l_{\textrm{gican}}(x) = E^i(\uvec x\times\uvec\nabla) A^i_\bot \quad \textrm{and} \quad \uvec s_{\textrm{gican}}(x) = \uvec E\times\uvec A_\bot. \eeq
  There are several reasons why $\uvec J_{\text{gican}}$, in spite of the above issues, is relevant and important in laser optics:\nl
 a) In general $\bm{L_{\textrm{gican}}}$ does not commute with $\bm{S_{\textrm{gican}}}$, but
\beq [L_{\textrm{gican},\,z} \, , S_{\textrm{gican},\,z}] =0  \eeq
so  $L_{\textrm{gican},\,z} \,\textrm{and}\, S_{\textrm{gican},\,z}$  can be measured simultaneously, even at a quantum level.\nl
b) Laser optical beams are almost invariably treated in the \emph{paraxial approximation}. Although the eigenvalues of $S_{\textrm{gican},\,z}$ and $L_{\textrm{gican},\,z}$  are continuous, in general,  for paraxial fields they are
approximately integer multiples of $\hbar $  \nl
c) For a paraxial photon  absorbed by an atom the photon's $S_{\textrm{gican},\,z}$ is transferred, approximately, to the internal AM of the atom and the $L_{\textrm{gican},\,z}$ approximately to the motion of the atom as a whole. \nl
Hence, for paraxial fields, $J_{\textrm{gican},\,z},\, L_{\textrm{gican},\,z}$ and $ S_{\textrm{gican},\,z}$   seem to function as perfectly good physical angular momenta. \nl
In complete contrast to all of the above, textbooks on classical electrodynamics teach us that the momentum density in a electromagnetic field is given by the poynting vector
\beq \label{Poynting} \bm{p}_{\textrm{poyn}}(x)=\textrm{poynting vector} =  \bm{E} \times \bm{B }  \eeq
and that the angular momentum density, which I shall call the  \emph{poynting version} (poyn), \footnote{This is called ``Belinfante"   by particle physicists. Poynting did not give this expression. I believe Belinfante was the first to do so.}  is  given by
\beq \label{bel}
\bm{j}_{ \,\text{\textrm{poyn}}}(x)= \bm{r }\times (\bm {E }\times \bm {B}).
   \eeq
   with total AM
 \beq \label{Jpoyn} \bm{J}_\text{poyn}= \int d^3x \,[\bm{r }\times (\bm {E }\times \bm {B})]. \eeq
Although this has the structure of an orbital AM, \emph{i.e.} $\bm{r}\times\bm{p}_{\textrm{poyn}}$, it is the \emph{total} photon angular momentum, and it is not split into orbital and spin parts.\nl
Now the integrands of Eqs.~(\ref{Jpoyn}) and (\ref{Jgican}) can be shown to differ by a divergence, so that
\beq  \bm{ J}_\text{poyn}= \bm{J}_\text{gican}  + \textrm{surface term}, \eeq
and \emph{if} the fields vanish at infinity the surface term vanishes so that
\beq \label{Jsequal} \bm{ J}_\text{poyn}= \bm{J}_\text{gican}.  \eeq
That is fine for classical fields, but quantum fields are operators, and it is extremely non-trivial to try  to attach meaning to the concept of operators vanishing at infinity. Hence, one must conclude, that as operators,
\beq \bm{J}_\text{poyn} \neq \bm{J}_{\text{gican}}.  \eeq
Let us return now to the consideration of classical paraxial optical beams. The key point is that even if Eq.~(\ref{Jsequal})  holds, i.e. even if  $\bm{ J}_\text{poyn}=\bm{ J}_\text{gican}$, their densities are different, and the intriguing question arises as to whether a laser optics measurement sensitive to the AM \emph{density} could decide which of the two densities, gauge invariant canonical or poynting, correctly describes the physical AM carried by the optical beam.\nl
Ever since the 1990s there have been beautiful laser optics experiments which measure the transfer of AM from the field to a particle. The early experiments \cite{PhysRevLett.75.826,PhysRevA.54.1593,Simpson:97} used particles  whose dimensions were comparable to the beam diameter and
hence were sensitive only to \emph{total} $\bm{J}$, and so could not distinguish between $\bm{J}_\text{poyn} \,\,\textrm{and} \,\, \bm{J}_{\text{gican}}$. Later experiments \cite{PhysRevLett.88.053601,PhysRevLett.91.093602} used very small particles and were able to record the motion of the particle as a function of distance $\rho$ from the beam axis, but were not sensitive enough to distinguish between the gauge invariant canonical and poynting densities. \nl
The general concept of these experiments is as follows:
 (a) A tiny particle is trapped in a ring of radius $\rho$ in, for example, a Bessel beam  \nl
 (b) The particle spins about its CM driven by the spin AM absorbed \nl
 (c) The particle rotates in the ring driven by the azimuthal force, which is proportional to the orbital AM of the beam \nl
(d) Because of viscous drag and torque there results  limiting angular velocities for the rotation  and the spin. \nl
Hence, in principle, the local orbital and spin densities can be measured as a function of $\rho$ if the particle is small enough and its position can be sufficiently accurately controlled. The key question is how different do we expect the densities to be? \nl
In what is regarded as the foundation paper on optical angular momentum, Allen et. al. \cite{PhysRevA.45.8185} utilized the \emph{poynting version} for the total AM and studied its structure in the paraxial approximation. The standard form for a monochromatic paraxial electric field propagating in the z-direction, is
\beq \bm{E}(\bm{r})=\Big(u(\bm{r}),\, v(\bm{r}),\, \frac{-i}{k}\big(\frac{\partial u}{\partial x} + \frac{\partial v}{\partial y}\big)\Big)e^{i(kz-\omega t)}  \eeq
where, choosing,
\beq v(\bm{r})=i\sigma \,u(\bm{r}) \quad \textrm{with} \quad\sigma=\pm 1  \eeq
 corresponds, approximately, to right  or left circular polarization.
Allen et.al. worked specifically with a Laguerre-Gaussian field, but one obtains the  same result for any vortex field
with an azimuthal mode index $l$, and with the form, in cylindrical coordinates, $(\rho, \, \phi, \, z)$,
 \beq  u(\rho,\phi,z)= f(\rho, z) e^{il\phi}.  \eeq
 In the following we shall indicate relations that are correct only in paraxial approximation by $\pareq$.
For the cycle average of the z-component of the poynting density  , $\langle j_{\textrm{poyn,\,z}} \rangle$, per unit power, modulo $\frac{\epsilon_0}{ \omega}$, Allen et. al. obtained
\beq \label{parpoyn} \langle j_{\textrm{poyn,\,z}} \rangle \, \pareq \, l|u|^2  - \frac{\sigma}{2}\rho\frac{\partial |u|^2 }{\partial\rho} \eeq
and interpreted the terms on the RHS as representing orbital and spin AM respectively. It should be stressed that this clean separation  into ``orbital" and ``spin" parts is only true in paraxial approximation. For a genuine Maxwell field there are terms in which $l$ and $\sigma$ are mixed together (See Section~3 of \cite{Allen1999291}).\nl
On the other hand, using the gauge invariant canonical version, one obtains
\beq \label{pargican} \langle j_{\textrm{gican,\,z}} \rangle \, \pareq \, l|u|^2 + \, \sigma|u|^2 .  \eeq
Here, on the basis of Eq.~(\ref{jgican}),  a clean separation into ``orbital" and ``spin" terms holds also for exact Maxwell fields, but the particular simple form Eq.~(\ref{pargican}) is valid only in paraxial approximation. Moreover, as discussed earlier, the terms in Eq.~(\ref{jgican}) function as physical AM only to the extent that the paraxial approximation is valid.  \nl
In summary, working always in paraxial approximation,  we have two competing expressions for the spin AM, the poynting and the gauge invariant canonical, and there is a clear difference between  them:
\beq \label{difference} \langle s_{\textrm{poyn,\,z}} \rangle \pareq - \frac{\sigma}{2}\rho\frac{\partial |u|^2 }{\partial\rho} \qquad \quad
   \langle s_{\textrm{gican,\,z}} \rangle\pareq   \sigma|u|^2   \eeq
and the challenging question is:  could an experiment decide which corresponds to the physical spin AM carried by an optical vortex beam? To study the feasibility of this, as an example,  we compare in Fig.~1. $\langle s_{\textrm{poyn,\,z}} \rangle $ and $\langle s_{\textrm{gican,\,z}} \rangle $, as function of $\rho$, for a $J_2(k_t\rho)$  Bessel beam.
\begin{figure}[H]
%%% in [  ] can have h (here), b (bottom),
%%t (top), p (new page)
\includegraphics[ width=0.5\textwidth]{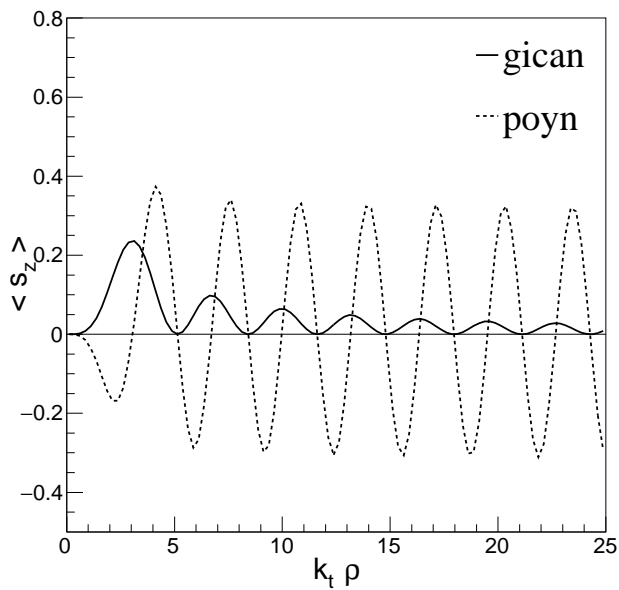}
%%% filename in {  };  in [  ]  can have:
%%width=  cm, height=  cm, angle = %%
\caption{Comparison of the $\rho$ dependence for the cycle average of the z-component of the poynting and gauge invariant canonical spin AM, for a $J_2(k_\bot \rho)$ Bessel beam. (Courtesy of Patrick Dunne) }
\end{figure}
Clearly there is a dramatic difference in the $\rho$-dependence of the two versions, and  this  should be measurable. Note, however, that integrated across a   ``bright ring"
\begin{equation} \int_{\textrm{ring}} d\rho \,\rho \, \langle s_{\textrm{poyn,\,z}} \rangle =\int_{\textrm{ring}} d\rho \,\rho \, \langle s_{\textrm{gican,\,z}} \rangle  \eeq
so that a successful measurement would require extremely small particles, i.e. with dimensions considerably smaller than the ring width.
 The situation for a Laguerre-Gaussian beam with radial mode index $p> 1$ is similar.\nl
 The behaviour of the poynting version in Fig.~1  looks,  intuitively,  unphysical, suggesting that the gican version is the physically relevant one. And, indeed,  there are reports in the literature of experiments which favour the gican version, but they are less direct than the type of experiment discussed above. For example, in an   unpublished paper in 2012, Chen and Chen \cite{XSChen} argue that the  Ghai et. al.  experiment in 2009 \cite{Ghai:2009} on the shift of diffraction fringes in the single slit diffraction of beams with a phase singularity favours the gican version. And other arguments in favour of the gican version can be found in the review of Bliokh and Nori  \cite{Bliokh:2015doa} and in \cite{Antognozzi, Bliokh:2014X}. \nl
 Ultimately, however,  a convincing demonstartion in favour of one or the other  requires an experiment of the type discussed above, which measures directly the transfer of spin AM from the beam to the particle.\nl
 
 I have benefited from interesting discussions with Konstantin Bliokh, Robert Boyd and Halina Rubinsztein-Dunlop, and am grateful to Patrick Dunne for help in preparing this paper.\nl
 \nl
 \nl

\bibliography{Elliot_General}% Produces the bibliography via BibTeX.

\end{document}